\documentclass[preprint,pteplogo]{ptephy}

\preprintnumber{KUNS-2698, YITP-17-93} 

\usepackage{amsmath} 
\usepackage{times}
\usepackage{graphicx,color}
\usepackage{multirow}

\usepackage[normalem]{ulem}  

\renewcommand\sout{\bgroup \color{red} \ULdepth=-.5ex \ULset}
\newcommand{\NTP}{{N_\mathrm{TP}}}

\newcommand{\NMC}{{N_\mathrm{MC}}}
\newcommand{\ND}{{N_\mathrm{D}}}

\newcommand{\fW}{f_{\scriptscriptstyle{W}}}
\newcommand{\fH}{f_{\scriptscriptstyle{H}}}

\newcommand{\SHW}{{S_\mathrm{HW}}}
\newcommand{\SHWPA}{{S_\mathrm{HW}^\mathrm{(PA)}}}
\newcommand{\beq}{\begin{eqnarray}}
\newcommand{\eeq}{\end{eqnarray}}
\mathchardef\mhyphen="2D 
\newcommand{\fm}{\mathrm{fm}}
\begin{document}

\title{Entropy production and isotropization in Yang-Mills theory with use of quantum distribution function}

\author{\name{Hidekazu Tsukiji}{1,\ast}, \name{Teiji Kunihiro}{2}, \name{Akira Ohnishi}{1}, and \name{Toru T. Takahashi}{3}}

\address{\affil{1}{Yukawa Institute for Theoretical Physics, Kyoto University,
Kyoto 606-8502, Japan}
\affil{2}{Department of Physics, Faculty of Science, Kyoto University,
Kyoto 606-8502, Japan}
\affil{3}{Gunma National College of Technology, 
Gunma 371-8530, Japan}
\email{tsukiji@yukawa.kyoto-u.ac.jp}}

\begin{abstract}
We investigate thermalization process in relativistic heavy ion collisions 
in terms of the Husimi-Wehrl (HW) entropy defined with
 the Husimi function, a quantum distribution function in a phase space.
We calculate the semiclassical time evolution of the HW entropy in Yang-Mills  field theory with the phenomenological initial field configuration known as the McLerran-Venugopalan model in a non-expanding geometry, which has instabilty triggered by initial field fluctuations.
HW-entropy production implies the thermalization of the system and it reflects the underlying dynamics such as chaoticity and instability.
By comparing the production rate with the Kolmogorov-Sina\"i rate, we find that the HW entropy production rate is significantly larger
than that expected from chaoticity.
We also show
that the HW entropy is finally saturated when the system reaches a quasi-stationary state.
The saturation time of the HW entropy is comparable with that of pressure isotropization, 
which is around $1$ fm/c in the present calculation in the non-expanding geometry.
\end{abstract}

\subjectindex{}

\maketitle

\section{Introduction}

A new form of matter consisting of deconfined quarks and gluons
is formed in high-energy heavy-ion collisions
at Relativistic Heavy Ion Collider (RHIC) 
and Large Hadron Collider (LHC) \cite{WPBRAHMS,WPPHENIX,WPPHOBOS,WPSTAR,Muller12}.
The created matter
is opaque for colored particles,
shows hydrodynamical behavior and collectivity of quarks,
and finally decays into hadrons.
Then it is considered to be a quark gluon plasma (QGP)~\cite{Kolb:2003dz,Gyulassy:2004zy}.
While quantitative studies on the QGP properties 
are in progress using hydrodynamical models combined with jet and hadronic transport,
its formation process is not yet clear.
For example, the early thermalization problem remains
as one of the serious problems in high-energy heavy-ion collisions \cite{Heinz02}.
Hydrodynamical-model analyses suggest that
the created matter becomes close to local equilibrium
at $\tau_\mathrm{th}=0.6\mhyphen1.0~\fm/c$ after the contact,
and this thermalization time is significantly shorter than 
the perturbative QCD estimate~\cite{Baier01,Baier02}.

In tackling the early thermalization problem,
the classical Yang-Mills field plays an important role.
The created matter in the initial stage is described well by the classical Yang-Mills field,
and is often called ``glasma'' \cite{Glasma}.
In the glasma,
both the color-electronic and -magnetic fields are parallel
to the collision axis,
the pressure is anisotropic,
and the anisotropy leads to instabilities triggered by initial field fluctuations~\cite{Arnold05,Romatschke06-1,Romatschke06-2,Berges08,Berges09,Berges12,Fukushima12,Tsutsui15,Tsutsui16}.
Fluctuations of classical fields may be regarded as particles,
then the glasma instability is expected to produce many particles and cause early thermalization.
Actually, recent studies \cite{Epelbaum13,Ruggieri14,Ruggieri15,Kurkela15}
have successfully shown early-time isotropization of the pressure
required by the hydrodynamical model analyses.
For the detailed understanding of the thermalization process, however, 
we need to evaluate the entropy of the system,
which is a very important key concept
that characterises thermalization.
In Ref.~\cite{EntropyReview}, the required amount of entropy 
produced in the glasma is estimated to be 3000 per rapidity
or the 55 \% of the total entropy.
Nevertheless, in many of previous works \cite{Arnold05,Romatschke06-1,Romatschke06-2,Berges08,Berges09,Berges12,Fukushima12,Epelbaum13,Ruggieri14,Ruggieri15,Kurkela15,Taya:2016ovo,Berges:2017eom},
the entropy production itself has not been discussed,
and the relation between the isotropization and thermalization remains unclear.

Entropy production in the classical Yang-Mills field theory has been discussed
based on the Kolmogorov-Sina\"i entropy production rate (KS rate)~\cite{Muller,Biro1993,kunihiro,Iida13}
and the Husimi-Wehrl entropy~\cite{Tsukiji2015,Tsukiji2016}:
Entropy of classical systems is obtained by the Wehrl entropy\cite{Wehrl1978,Wehrl1979},
$S_\mathrm{W}=-\mathrm{Tr} f \log f$,
where $f$ is the phase space distribution function
and $\mathrm{Tr}$ denotes the integral over the phase space.
In quantum systems, the Wigner function~\cite{Wigner,Hillery:1983ms,Mrowczynski:1994nf,Fukushima:2006ax} ($f_W$)
is a candidate of the distribution function,
since it is defined through a mere Weyl transformation~\cite{Weyl1931} of the density matrix and should 
contain full information equivalent to the density matrix.
However the Wigner function is not appropriate as the distribution function
to discuss entropy production;
it is not semi-positive definite and cannot be regarded
as the phase space probability distribution.
In addition, even if $f_W \geq 0$ is satisfied everywhere,
the Wehrl entropy does not increase
in the semiclassical time evolution due to the Liouville theorem.

One possible solution for the phase space distribution function in calculating the Wehrl entropy 
is the Husimi function~\cite{Husimi} ($f_H$).
The Husimi function is obtained from the Wigner function by smearing
 in the phase space
within the allowance of the uncertainty principle, and it is shown to be semi-positive definite.
In fact, Husimi function is an expectation value of the density matrix with respect to 
the wave packet with the minimal uncertainty, which is nothing but a coherent state 
\cite{KTakahashi1986,KTakahashi1989}.
We call the Wehrl entropy defined with the Husimi function
the Husimi-Wherl (HW) entropy~\cite{KMOS,Tsukiji2015,Tsukiji2016,Wehrl1978,Wehrl1979}.
The HW entropy is shown to be approximately the same
as the von Neumann entropy at high temperatures~\cite{KMOS}.
In inverse harmonic oscillators,
the HW entropy is found to increase in time
and the growth rate agrees with the KS rate,
the sum of the positive Lyapunov exponents~\cite{KMOS}.
The increase of the HW entropy implies information loss
caused by instabilities and/or chaoticities combined
with the coarse-graining in the phase space,
and it is expected to play a crucial role in thermalization.
%

We can obtain the HW entropy in field theories
by regarding the field strength and its canonical conjugate momentum
as the phase space variables.
In Ref.~\cite{Tsukiji2016},
the present authors have calculated the semiclassical time evolution
of the HW entropy of the classical Yang-Mills fields with a random initial condition,
and have confirmed that the HW entropy growth rate
is consistent with the KS rate~\cite{kunihiro}.
This agreement suggests the entropy production is caused by the chaoticity
of the classical Yang-Mills fields,
since the KS rate characterizes the chaoticity of the system.

In this article, we discuss entropy production in Yang-Mills field theory
starting from the glasma-like configuration
given by the McLerran-Venugapalan(MV) model~\cite{MV1994,Kovner1995}
in the non-expanding geometry \cite{Iida14}
based on the framework developed in \cite{Tsukiji2015,Tsukiji2016}.
Quantum fluctuations are incorporated around the initial glasma-like field configuration,
and we compare the time scales of
the entropy production with that of other quantities such as the pressure isotropization and the equilibration of
the local energy distribution.

This paper is organized as follows.
In Sec.~\ref{Sec:Framework},
we introduce the quantum distribution functions and entropy
in field theories as well as the initial condition in the MV model
in the non-expanding geometry.
In Sec.~\ref{Sec:Method},
we explain the numerical method to calculate the semiclassical time evolution
of the HW entropy and pressure.
We show the results in the Sec.~\ref{Sec:Results}.
Section~\ref{Sec:Summary} is devoted to the summary of our work.

\section{Husimi-Wehrl entropy from classical Yang-Mills dynamics}
\label{Sec:Framework}

\subsection{Quantum distribution functions and entropy in Yang-Mills theory}

The Husimi-Wehrl entropy of the Yang-Mills field is obtained
as a natural extension of that in quantum mechanics
by regarding $(A(x),E(x))$ as canonical variables.
We define the Wigner and Husimi functions on the lattice,
as a straightforward extension of those in quantum mechanics~\cite{Mrowczynski:1994nf,Fukushima:2006ax}.
The semiclassical time-evolution of the Wigner function
is given by the classical equation of motion (see Eq.(\ref{EOM}) below),
then we can obtain the Husimi function
from thus constructed Wigner function at each time.

In the $\text{SU}(N_c)$ Yang-Mills field theory on a $L^3$ lattice
in the temporal gauge, 
the Hamiltonian in the non-compact formalism 
is given by
\beq
H=\frac{1}{2}\sum_{x,a,i} E^a_i(x)^2 +\frac{1}{4}\sum_{x,a,i,j} F^a_{i j}(x)^2,
\eeq
where $(A^{a i}(x), E^{a i}(x)=F^{a i 0}(x))$ are the canonical variables,
$F^a_{i j}=\partial_i A^a_j(x)-\partial_j A^a_i(x)+\sum_{b,c}f^{a b c}A^b_i(x)A^c_j(x)$ is the field strength tensor,
and $N_D=3L^3(N_c^2-1)$ is the total degrees of freedom (DOF).
We take the dimensionless gauge field $A$ and conjugate momentum $E$
and space-time variables $x$
normalized by the lattice spacing $a$ throughout this article.
Then the Wigner function $\fW[A,E;t]$ 
 is defined
by a Weyl transform of the density matrix $\hat{\rho}$ as
\begin{align}
\fW[A,E;t]=&\int \frac{DA'}{g}\, e^{i {E\cdot A'}/{\hbar g^2}}
\left\langle A+A'/{2}\mid\hat{\rho}(t)\mid{A-A'/2}\right\rangle,\label{fW}
\end{align}
where
$A\cdot E=\sum_{i,a,x} A^a_i(x) E^a_i(x)$ denotes the inner product.
It should be noted that
the coupling constant $g$ appears in the denominator in the integral measure,
since $g$ is included in the definitions of $A$ and $E$.
The expectation value of a physical quantity $X$ is given
by integrating the product of $\fW[A,E;t]$ and 
$X$ in the Weyl representation denoted by $X_{\rm W}[A,E]$ as;
\beq
\langle X\rangle(t)=\int D\Gamma\,\fW[A,E;t]\,X_{\rm W}[A,E]\ ,
\label{Eq:Expectation}
\eeq
where $D\Gamma=DA\,DE/(2\pi\hbar g^2)^{N_D}$.
For instance, let $X$ be transverse (longitudinal) pressure $P_{T,L}$.
The pressure is given by the diagonal part of the energy-momentum tensor $T^{\mu\nu}$.
The expectation value is then given by  
\beq
\langle P_{T}\rangle&=&\frac{1}{2}\langle T^{11}+T^{22}\rangle=\frac{1}{2} \langle E^{a 3}E^{a 3}\rangle+\frac{1}{2} \langle B^{a 3}B^{a 3}\rangle,\\
\langle P_L\rangle&=&\langle T^{33} \rangle=\langle E^{a}_{\perp}E^{a}_{\perp}\rangle+\langle B^{a}_{\perp}B^{a}_{\perp}\rangle-\frac{1}{2}\langle E^{a 3}E^{a 3}\rangle-\frac{1}{2}\langle B^{a 3}B^{a 3}\rangle,
\eeq
where the $E^{a}_{\perp} (B^{a}_{\perp})$ is the transverse component
of the color electric (magnetic) field, and
$E^{a}_{\perp}E^{a}_{\perp}=\frac{1}{2}E^{a 1}E^{a 1}+\frac{1}{2}E^{a 2}E^{a 2}$
($B^{a}_{\perp}B^{a}_{\perp}=\frac{1}{2}B^{a 1}B^{a 1}+\frac{1}{2}B^{a 2}B^{a 2}$).
The color magnetic field is defined as
$B^{a i}=-\frac{\epsilon^{i j k}}{2} F^{a}_{j k}$,
and the $\epsilon^{i j k}$ is a completely antisymmetric (Levi-Civita) tensor ($\epsilon^{123}=1$).
The time evolution of the Wigner function is derived
from the von Neumann equation,
\beq
\frac{\partial}{\partial t}\fW[A,E;t]
=\frac{\partial H}{\partial A}\cdot\frac{\partial \fW}{\partial E}
-\frac{\partial H}{\partial E}\cdot\frac{\partial \fW}{\partial A}
+\mathcal{O}(\hbar^2)
.\label{EOM}
\eeq
In the semiclassical approximation in which
we ignore $\mathcal{O}(\hbar^2)$ terms,
$\fW$ is found to be constant along the classical trajectory
given by the classical equation of motion (EOM) \cite{Polkovnikov:2009ys},
\beq
\dot{E}=-\frac{\partial H}{\partial A}, \, \dot{A}=\frac{\partial H}{\partial E}.\label{cEOM}
\eeq

The Husimi function is defined as the smeared Wigner function
with the minimal Gaussian packet,
\begin{align}
\fH[A,E;t]=&\int 
D\Gamma'\ G(A-A',E-E';\Delta)
\fW[A',E';t]\ ,\label{fH}
\\
G(A,E;\Delta)=&
2^{N_D} \exp(-\Delta A^2/\hbar g^2-E^2/\Delta \hbar g^2)
\ ,\label{Eq:Gaussian}
\end{align}
where $\Delta=a\Delta_\mathrm{phys}$ is the dimensionless 
parameter corresponding to the Gaussian-smearing range.
It should be noted that the Husimi function is also obtained
as the expectation value of the density matrix in the coherent state
as in quantum mechanics~\cite{KTakahashi1986}.
Then the Husimi function is semi-positive definite, $f_H[A,E;t]\ge0$,
while the Wigner function is not. 
We finally define the Husimi-Wehrl entropy 
as the Boltzmann's entropy or the Wehrl's classical entropy~\cite{Wehrl1978}
by adopting the Husimi function for the phase space distribution,
\beq
\SHW(t)=-\int D\Gamma\,\fH[A,E;t]\,\log \fH[A,E;t].\label{SHW}
\eeq
The HW entropy is gauge invariant,
and the semiclassical time evolution does not break the gauge invariance
as shown in Appendix A.

\subsection{Initial condition}

We consider two nuclei moving at the velocity of light along the $z$ axis.
These nuclei collide at time $t=0$ and $z=0$,
and glasma is formed between the two nuclei.
In the framework of the color glass condensate (CGC),
the gluons with small Bjorken $x$ are described by the classical field
and those with large Bjorken $x$ and quarks are regarded as color sources.
The color-source distribution is assumed to be Gaussian
in the McLerran-Venugopalan (MV) model \cite{MV1994,Kovner1995}.

We adopt a glasma-like initial condition
which mimics the MV model in the non-expanding geometry~\cite{Iida14}.
As in the MV model,
we generate the Gaussian random color sources for a target nucleus $\rho^{(t)}$
and a projectile $\rho^{(p)}$,
\begin{align}
\langle \rho^{(t)a}({\bf x_\perp}) \rho^{(t)b}({\bf y_\perp}) \rangle
=g^4\mu_\mathrm{phys}^2\delta^{ab}
\delta^{(2)}({\bf x_\perp}-{\bf y_\perp}),\nonumber\\
\langle \rho^{(p)a}({\bf x_\perp}) \rho^{(p)b}({\bf y_\perp}) \rangle
=g^4\mu_\mathrm{phys}^2\delta^{ab}
\delta^{(2)}({\bf x}_\perp-{\bf y}_\perp), 
\label{Eq:MV}
\end{align}
where ${\bf x}_\perp\equiv (x,y)$ and $a, b$ are the color indices.   
On the lattice, the delta function
$\delta^{(2)}({\bf x}_\perp-{\bf y}_\perp)$ is replaced by 
the Kronecker delta
$\delta_{{\bf x}_\perp,{\bf y}_\perp}/a^2$,
and Eq.~\eqref{Eq:MV} reads
$\langle \rho^{(i)a}({\bf x_\perp}) \rho^{(j)b}({\bf y_\perp}) \rangle
=g^4\mu^2\delta^{ij}\delta^{ab}\delta_{{\bf x}_\perp,{\bf y}_\perp} (i,j=p,t)$,
where $\mu=a\mu_\mathrm{phys}$ and $\rho^{(i)a}$ is given in the lattice unit.
%
Gauge fields are given by $\alpha_i^{(t)}=iV\partial_i V^\dagger$ and 
$\alpha_i^{(p)}=iW\partial_i W^\dagger$ ($i=x,y$) with Wilson lines,
$V^\dagger({\bf x_\perp})=e^{i\Lambda^{(t)}({\bf x_\perp})}$ and 
$W^\dagger({\bf x_\perp})=e^{i\Lambda^{(p)}({\bf x_\perp})}$,
which are created by the color sources;
\begin{align}
-{\bf \partial}^2_\perp\Lambda^{(t)}({\bf x_\perp})=\rho^{(t)}({\bf x_\perp}),
\quad
-{\bf \partial}^2_\perp\Lambda^{(p)}({\bf x_\perp})=\rho^{(p)} ({\bf x_\perp}).
\end{align}
Gauge fields, electric fields and magnetic fields are then given by
\begin{align}
& A^i=\alpha_i^{(t)} + \alpha_i^{(p)}, A^z=0,
\label{Eq:MVA}\\
& E^i=0, E^z 
=i\sum_i\left(\left[\alpha_i^{(t)},\alpha_i^{(p)}\right]\right),
\label{Eq:MVE}\\
& B^i=0, B^z=i\left(\left[\alpha_1^{(t)},\alpha_2^{(p)}\right]+\left[\alpha_1^{(p)},\alpha_2^{(t)}\right]\right).
\label{Eq:MVB}
\end{align}


The above gauge fields are classical and uniform in the $z$ direction,
and there is no quantum fluctuations for a given source.
In order to make the initial Wigner function $f_{W}[A,E,t=0]$
taking into account quantum fluctuations,
the uncertainty relation between $A$ and $E$,
the initial Wigner function is set to be
a glasma-like field configuration with a Gaussian fluctuation around it.
With $A_{\rm MV}$ and $E_{\rm MV}$ being the solutions of
Eqs.~\eqref{Eq:MVA} and \eqref{Eq:MVE},
the initial Wigner function is obtained as
\beq
f_W[A,E;t=0]=G(A-A_{\rm MV},E-E_{\rm MV};\omega)
\ ,
\label{Initial}
\eeq
where $\omega=a \omega_\mathrm{phys}$ is the parameter of the Gaussian width.

\subsection{Physical scale}

We have two dimensionful parameters,
$g^2\mu_\mathrm{phys}$ and $\omega_\mathrm{phys}$,
in the initial condition,
one dimensionful parameter, $\Delta_\mathrm{phys}$,
in the calculation of the HW entropy,
and one dimensionless parameter, $\hbar g^2$,
in addition to the lattice spacing $a$.
The factor $\hbar g^2$ appears from the field redefinition,
$gA\to A$ and $gE \to E$,
then the uncertainty relation is modified as 
$(\Delta A)^2\,(\Delta E)^2 \geq (\hbar g^2/2)^2$
for each component of $A$ and $E$.
This relation is consistent with the {\em classical} field dominance
in the weak coupling regime.
Since the saturation scale $Q_s$ is the fundamental scale
in the color glass condensate,
we take $\mu_\mathrm{phys}\simeq Q_s$ and $\omega_\mathrm{phys}\simeq Q_s$.

We now set the physical scale.
We consider heavy-ion collisions at RHIC and LHC energies,
then $g=1 (\alpha_s=0.15)$ and $Q_s\simeq 2~\mathrm{GeV}$ may be reasonable.
We also assume that the total lattice area in the $xy$ plane
is equal to the transverse area of the colliding nuclei.
Then the parameters are fixed as
\beq
\mu_\mathrm{phys}&\simeq& Q_s\simeq2\ {\rm[GeV]}\ ,\\
a L&\simeq&\sqrt{\pi}R_A\simeq7\sqrt{\pi}\ {\rm[fm]}\ .
\eeq
From these equations, we get
\beq
g^2\mu_\mathrm{phys} aL\simeq120.\label{scale}
\eeq
The lattice spacing $a$ is inversely proportional to the lattice size $L$.

\section{Numerical methods}\label{Sec:Method}

It is not an easy task to perform numerical calculation of the HW entropy,
Eq.~\eqref{SHW}, especially in field theories.
We need $2N_D$ dimensional integral of a function $f_H$,
which additionally requires $2N_D$ dimensional integral to obtain,
where $N_D$ is very large in field theories.
The logarithmic term $-\log \fH$ takes a large value when $\fH$ is small
and the integrand exhibits an acute peak.
The Monte-Carlo method is then effective and necessary
for a large-dimensional integral.

We have developed numerical methods
to calculate the time evolution of the Husimi-Wehrl entropy
in semiclassical approximation in quantum mechanical systems~\cite{Tsukiji2015}
and the Yang-Mills field theory~\cite{Tsukiji2016}.
In this section, we recapitulate our formalism.
We introduce two methods based on the test particle (TP) method
to calculate the HW entropy,
which were applied to Yang-Mills field theory in Ref.~\cite{Tsukiji2016}.
The test particle method is applied also to calculate other physical quantities such as pressure.

\subsection{Test particle method and Husimi-Wehrl entropy}

In the TP method,
we express the Wigner function by a sum of the delta functions,
\begin{align}
\fW[A,E;t]=&\frac{(2\pi\hbar g^2)^\ND}{N_\mathrm{TP}}
\sum_{\alpha=1}^{N_\mathrm{TP}}
\delta^\ND(A-A_\alpha(t))\,\delta^\ND(E-E_\alpha(t)),\label{Eq:WignerTP}
\end{align}
where $N_{\rm TP}$ is the total number of the test particles,
the number of the delta functions used to express the Wigner function.
The variables
$(A_\alpha(t),E_\alpha(t))=\{(A_{\alpha,i}^a(\bold{x},t),E_{\alpha,i}^a(\bold{x},t)) \mid i=1,2,3, a=1,2,\ldots N_c^2-1\}$
represent the phase space coordinates of test particles
at time $t$.
The initial conditions of the test particles,
$(A_\alpha(0),\,E_\alpha(0))$\, $(i=1,\,2,\dots,\,\NTP)$,
are chosen so as to well sample $\fW[A, E, 0]$ in Eq. (\ref{Initial}).
The time evolution of the coordinates $(A_\alpha(t),E_\alpha(t))$ 
is determined by the canonical equation of motion, Eq.~(\ref{cEOM}),
which is derived from the EOM for $\fW[A, E, t]$
in the semiclassical approximation.
Substituting the test-particle representation 
of the Wigner function Eq.~\eqref{Eq:WignerTP} into Eq.~\eqref{fH},
the Husimi function is readily expressed as 
\begin{align}
\fH[A,E;t]=&\frac{1}{\NTP} \sum_{\alpha=1}^{\NTP} 
G(A-A_\alpha(t),E-E_\alpha(t);\Delta)
\ .\label{Eq:HusimiTP}
\end{align}
It is noteworthy that the 
Husimi function here is a smooth function
in contrast to the corresponding Wigner function in Eq.~\eqref{Eq:WignerTP}.

With the Wigner function Eq.\eqref{Eq:HusimiTP},
the HW entropy in the test-particle method Eq.~\eqref{SHW} is now obtained as,
\begin{align}
S_\mathrm{HW}^\mathrm{(TP,pTP)}
&=-\frac{1}{\NTP}
\sum_{\alpha=1}^\NTP
\int D\Gamma\,G(A\!-\!A_\alpha,E\!-\!E_\alpha;\Delta)
\log\left[\frac{1}{\NTP} \sum_{\beta=1}^{\NTP} 
G(A\!-\!A_\beta,E\!-\!E_\beta;\Delta)
\right]
\nonumber\\
&\simeq-
\frac{1}{\NMC\NTP}\sum_{k=1}^\NMC\sum_{\alpha=1}^\NTP
\log\left[
\frac{1}{\NTP} \sum_{\beta=1}^{\NTP} 
G(A_\alpha\!-\!A_\beta\!+\!\mathcal{A}_k,E_\alpha\!-\!E_\beta\!+\!\mathcal{E}_k;\Delta)
\right].\label{Eq:SHW2}
\end{align}
Note here that the integral over $(A,E)$ has a support 
only around the positions of 
the test particles $(A_\alpha(t),\,E_\alpha(t))$ 
due to the Gaussian function for each $\alpha$,
and we can effectively perform the Monte-Carlo integration.
We generate random numbers
$(\mathcal{A}_{\alpha, k},\mathcal{E}_{\alpha, k})$ $(k=1,\cdots,\NMC)$
with zero mean and standard deviations of $(\sqrt{\hbar g^2/2\Delta},\sqrt{\hbar g^2\Delta/2})$,
with $\NMC$ being the total number of Monte-Carlo samples.
Then we obtain the HW entropy as shown in the second line of Eq.~(\ref{Eq:SHW2}).
The width parameter $\Delta$ needed to define the Husimi function
$\fH$ is set to be $\Delta/\omega=1$.
At present, $\Delta$ is treated merely as an input parameter.
We have checked
the dependence of results on $\Delta$ and confirm that
main conclusions remain unchanged.

The TP method has a following problem.
In the case where $\alpha=\beta$ in Eq.~\eqref{Eq:SHW2},
the Husimi function, the argument of the logarithm,
tends to take a large value, which generally leads to an underestimate of the HW entropy.
Since this underestimate arises
from the $\fW$ sampled with a finite number of delta functions (test particles),
the HW entropy in the TP method is essentially underestimated,
though this artifact vanishes when $N_{TP}\rightarrow \infty$.
In order to evade the problem,
we also introduce a parallel test particle (pTP) method,
where we prepare independent sets of test particles
$(A_\alpha,E_\alpha)$ and $(A_\beta,E_\beta)$
for in and out of the logarithm in Eq.~\eqref{Eq:SHW2}.
In the pTP method,
the HW entropy tends to be overestimated.
The phase space distance of test particles grows exponentially
in chaotic or unstable systems,
then we may not have any test particle $(A_\beta,E_\beta)$ inside the logarithm
in the vicinity of the test particle $(A_\alpha,E_\alpha)$
prepared outside the logarithm.
In this case, the argument of the logarithm becomes very small,
and $-\log \fH$ is overestimated.
While both the TP and pTP methods have 
problems stemming from the formalism,
the results should converge at large $\NTP$ from below and above
in the TP and pTP methods, respectively,
and the converged value of $\SHW$
exists between the TP and pTP results at a finite $\NTP$.

\subsection{Product ansatz}

While the TP and pTP methods can be, in principle, applied
to the field theory on the lattice,
the DOF is large and numerical-cost is demanding.
For example, we need to adopt very large number of test particles, $\NTP$,
to make the Monte-Carlo integration converge.
Since the Husimi function is equivalent to the expectation value
of the density matrix
in the coherent state, it has a value in the range of $0\leq \fH \leq 1$.
The Gaussian Eq.~\eqref{Eq:Gaussian} take the maximal value $2^{N_D}$,
then the required number of test particles is $\NTP > 2^{N_D}$
in order to respect the $\fH$ range.
Thus we need to invoke some approximation scheme in practical calculations.

We here adopt a product ansatz to avoid this difficulty.
In the ansatz, we assume that the total Husimi function is given
as a product of that for each degree of freedom,
\beq
\fH^\mathrm{(PA)}[A,E;t]=\prod^{N_D}_I \fH^{(I)} (A_I,E_I;t)\ ,
\label{Eq:PA}
\eeq
where
$I=(i,a)$ denotes the direction ($i=x,y,z$) and color indices ($a=1,2,3$),
and $\fH^{(I)}=\int \prod_{J\not=I} dA_J dE_J/2\pi\hbar g^2\, \fH[A,E;t]$
is obtained by integrating out other degrees of freedom than $I$.
By substituting this ansatz into Eq.~\eqref{SHW},
we obtain the HW entropy as a sum of the HW entropy 
for each
degree of freedom;
\beq
\SHWPA
=\sum^{N_D}_{I=1}S_{\rm HW}^{(I)}
=-\sum^{N_D}_{I=1}\int\frac{dA_I dE_I}{2\pi \hbar g^2}
\fH^{(I)}\,\log \fH^{(I)}\ .\label{SHWPA}
\eeq

Some comments are in order here;
First, the entropy in the product ansatz $S_\mathrm{HW}^\mathrm{(PA)}$
gives the upper bound of $S_\mathrm{HW}$ due to the subadditivity of entropy~\cite{Tsukiji2016};
\begin{align}
S_\mathrm{HW}
\leq S_\mathrm{HW}^\mathrm{(PA)}.
\end{align}
It is found that the HW entropy obtained with product ansatz is found to 
overestimate the entropy by 10-20 \%
in a few-dimensional quantum mechanical system~\cite{Tsukiji2016}.
Secondly, the maximum value of the HW entropy in the TP method is shifted
with the product ansatz,
while the minimum value remains unchanged.
For a one-dimensional case,
there is a minimum of $\SHW=1$~\cite{Wehrl1979,Lieb1978}.
When the Wigner function is a Gaussian, $\fW(A,E)=G(A,E;\omega)$,
the Husimi function is also a Gaussian,
$\fH(A,E)=[2\sqrt{\Delta\omega}/(\Delta+\omega)]^{N_D}
\exp[-(\Delta\omega\,A^2+E^2)/\hbar g^2(\Delta+\omega)]$,
and the HW entropy is found to be
$\SHW=N_D(1-\log[2\sqrt{\Delta\omega}/(\Delta+\omega)]) \geq N_D$.
The equality holds when we take $\Delta=\omega$.
The HW entropy will have an upper bound in the TP method,
when all the test particles are separated from each other,
and we find
$\SHW \leq N_D+\log(\NTP/2^{N_D})$~\cite{Tsukiji2015}.
In the TP method with the product ansatz,
the HW entropy for each DOF has the above upper bound for $N_D=1$,
$S_\mathrm{HW}^{(I)} \leq 1+\log(\NTP/2)$.
Thus the upper bound of the HW entropy with the product ansatz
becomes larger than that without the ansatz,
\begin{align}
S_\mathrm{HW}^\mathrm{(PA)} \leq N_D\left[1+\log(\NTP/2)\right]
\ .
\end{align}
Thirdly, the HW entropy in the product ansatz $S_\mathrm{HW}^\mathrm{(PA)}$
is not gauge invariant.
Nevertheless we might expect that the gauge dependence does not cause serious problems in entropy production
because gauge degrees of freedom dose not significantly contribute
to chaoticity and instability~\cite{Iida13,Tsutsui15},
and that the production rate of the HW entropy from random initial condition
in the product ansatz agrees with the gauge invariant KS rate~\cite{Tsukiji2016}.

\subsection{Vacuum subtraction}

When we calculate observables in field theories,
it is generally necessary to subtract vacuum expectation values.
It also applies to the present semiclassical treatment.
Let $X$ be a physical quantity and $\langle X\rangle_{\rm MV}$ be the expectation value calculated by using the Winger function,
as given in Eqs.~\eqref{Eq:Expectation} and \eqref{Eq:WignerTP}.
When we calculate an expectation value of $X$,
we subtract the vacuum contribution $\langle X\rangle_{\rm vac}$
arising from quantum fluctuations.
We have evaluated the vacuum expectation value
by using the fluctuation part of $(A,E)$,
\begin{align}
\langle X(t)\rangle
=&\langle X(t)\rangle_\mathrm{MV}-\langle X(t=0)\rangle_{\rm vac}\nonumber\\
=&\frac{1}{N_\mathrm{TP}}\sum_{\alpha=1}^{N_\mathrm{TP}}
\left[
X(A_\alpha(t),E_\alpha(t))-X(\delta A_\alpha(0),\delta E_\alpha(0))
\right]\ .
\end{align}
with $A_\alpha(0)=A_{\rm MV}+\delta A_\alpha(0)$
and  $E_\alpha(0)=E_{\rm MV}+\delta E_\alpha(0)$.

For example, in the case of $X=P_{T,L}$,
the expectation values are given by
\begin{align}
\langle P_{T}(t)\rangle
=&\frac{1}{2} \langle E^{a 3}(t)E^{a 3}(t)\rangle
 +\frac{1}{2} \langle B^{a 3}(t)B^{a 3}(t)\rangle
\nonumber\\
-&\left[\frac{1}{2} \langle \delta E^{a 3}(0)\delta E^{a 3}(0)\rangle
 +\frac{1}{2} \langle \delta B^{a 3}(0)\delta B^{a 3}(0)\rangle\right],\\
\langle P_L(t)\rangle
=&\langle E^{a}_{\perp}(t)E^{a}_{\perp}(t)\rangle
 +\langle B^{a}_{\perp}(t)B^{a}_{\perp}(t)\rangle
 -\frac{1}{2}\langle E^{a 3}(t)E^{a 3}(t)\rangle
 -\frac{1}{2}\langle B^{a 3}(t)B^{a 3}(t)\rangle\nonumber\\
-&\left[\langle \delta E^{a}_{\perp}(0)\delta E^{a}_{\perp}(0)\rangle
 +\langle \delta B^{a}_{\perp}(0)\delta B^{a}_{\perp}(0)\rangle
 -\frac{1}{2}\langle \delta E^{a 3}(0)\delta E^{a 3}(0)\rangle
 -\frac{1}{2}\langle \delta B^{a 3}(0)\delta B^{a 3}(0)\rangle\right]
\ .\nonumber\\
\end{align}

\section{Results}\label{Sec:Results}

We shall now discuss the numerical results
of the time evolution of the HW entropy and the pressure
based on the numerical methods explained in Sec.~\ref{Sec:Method}.
We mainly show the results on the $64^3$ lattice,
and also show some of the results on the $16^3$ and $32^3$ lattices
for comparison.
The $64^3$ lattice may be a reasonable choice
to discuss heavy-ion collisions at RHIC and LHC
based on the classical Yang-Mills fields.
The classical Yang-Mills field theory is a low-energy effective theory
and has a ultraviolet cut off.
At $L=64$, the lattice spacing is $a\simeq 2Q_s^{-1}$,
which corresponds to the diameter of one color flux tube.

\subsection{Husimi-Wehrl entropy production}

\begin{figure}[tbhp]
\begin{center}
\includegraphics[width=100mm]{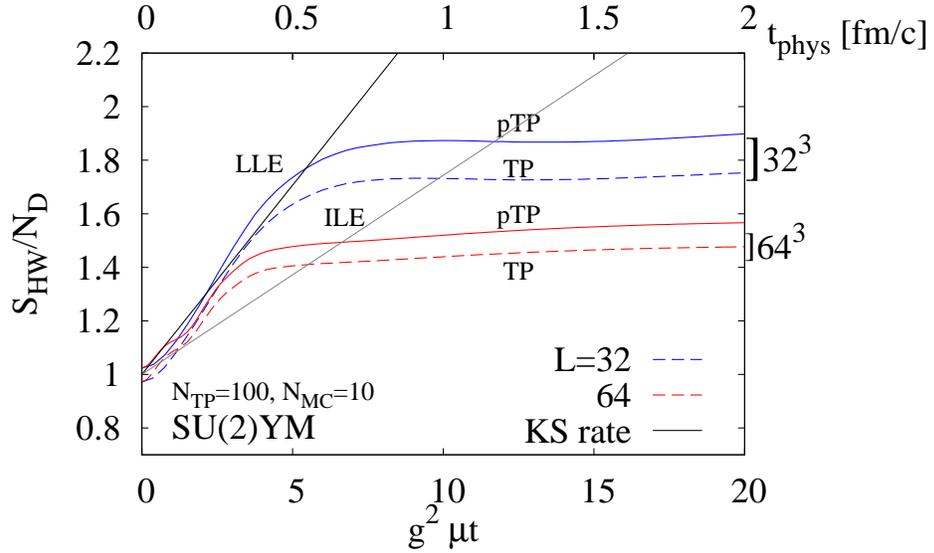}
\caption{The time evolution of HW entropy by TP and pTP methods in the SU(2) Yang-Mills (YM) lattice field theory on the $32^3$ and $64^3$ lattices.
The horizontal axis 
shows the time, where
the lower (upper) scale is the dimensionless time 
$g^2 \mu t$ (the physical time $t_\mathrm{phys}$ [fm/c]).
The vertical axis is the HW entropy per DOF, $S_\mathrm{HW}^\mathrm{(PA)}/N_D$.
The upper (lower) two lines are results on the $32^3$ ($64^3$) lattice.
In the both lattice results, the upper solid (lower dotted) line is the result
in the pTP (TP) method. 
The black (gray) straight line, ``KS rate'', is the entropy production rate given by the sum of the positive LLE (ILE) given in Appendix \ref{App:B}.}
\label{fig:HWE}
\end{center}
\end{figure}

In Fig.~\ref{fig:HWE}, we show the time evolution of the HW entropy
on the $32^3$ and $64^3$ lattices
obtained by the TP and pTP methods with the product ansatz.
The HW entropy per DOF starts from the minimum value, $\SHW/N_D=1$,
then increases rapidly and almost linearly until $g^2\mu t=3$ at 
almost a common rate on the $32^3$ and $64^3$ lattices,
and shows slow increase in the later stage.
In the later stage, e.g. $g^2\mu t=10$, the HW entropy takes a smaller
value on the $64^3$ lattice.
The pTP method gives the upper bound of the HW entropy 
and the TP methods gives the lower bounds,
then we can guess that the converged value in the limit of $\NTP \to \infty$
exists between the results of the two methods
as discussed in Ref.~\cite{Tsukiji2016}.

The growth rate of the HW entropy 
in the linearly increasing stage
may be
characterised by the Kolmogorov-Sina\"i (KS) rate,
which is the sum of positive Lyapunov exponents and reflects the underlying dynamics.
Due to the scale invariance of classical Yang-Mills,
the KS rate scales as
$\lambda_{\rm KS}/L^3=c_{\rm KS}\times\varepsilon^{1/4}$~\cite{kunihiro},
where $\varepsilon=\langle H\rangle/L^3$ is the energy density.
Then the HW entropy is expected to increase as
\begin{align}
\frac{\SHW(t)}{\ND}
=\frac{\SHW(t=0)}{\ND}+\frac{\lambda_{\rm KS}}{\ND}\,t
=1+\frac{c_{\rm KS}}{3(N^2_{\rm c}-1)}\,\frac{\varepsilon^{1/4}}{g^2\mu}\times
g^2\mu t.
\end{align}
We consider two types of the KS rate, 
the local and intermediate KS rates,
$\lambda^{\rm LLE}_{\rm KS}$ and $\lambda^{\rm ILE}_{\rm KS}$,
obtained from the local and intermediate Lyapunov exponents, LLE and ILE,
defined locally in time and in an intermediate time period,
respectively~\cite{kunihiro}.
For the SU(2) Yang-Mills theory, the coefficient is obtained as
$c^{\rm LLE (ILE)}_{\rm KS} \simeq 1.9 (1.0)$
by fitting to the data as shown in Appendix \ref{App:B}.

In the case of random initial condition discussed in
Appendix \ref{App:B},
the growth rate 
in the early time is characterized well by the local
KS rate, $\lambda_{\rm KS}^{\rm LLE}$.
On the other hand, the entropy growth rate in the intermediate time
toward the saturation agrees with the intermediate KS rate, $\lambda_{\rm KS}^{\rm ILE}$.
While the local KS rate obtained from the second derivative of the Hamiltonian at initial time is sensitive to the gluon-field configuration itself, the intermediate KS rate represents the intrinsic property of the chaotic system that dose not depend on initial conditions.

In Fig.~\ref{fig:HWE}, we compare the numerically obtained HW entropy
and that expected from the KS rates.
The black straight lines in Fig.~\ref{fig:HWE}
show the entropy increase expected
from the local and intermediate KS rates.
In the present calculation on the $64^3$ lattice,
the total energy (energy density) amounts to
$\langle{H}\rangle=6.5\times 10^5$ ($\varepsilon=2.48$),
then the slope from the local (intermediate) KS rates, 
$\frac{c^{\rm LLE(ILE)}_{\rm KS}}{3(N^2_{\rm c}-1)}\,\frac{\varepsilon^{1/4}}{g^2\mu}$,
is evaluated to be $0.14 (0.074)$.
As seen in Fig.~\ref{fig:HWE}, the growth rate
of the HW entropy in the early time 
is around $d\SHW/d(g^2\mu t)/\ND\simeq 0.14$,
which is close to the local KS rate and significantly larger than the
intermediate KS rate.
This comparison implies that we cannot explain the entropy production 
from the MV model initial condition only by intrinsic chaoticity, 
and that some instability may be the trigger of the entropy production.
In fact, the initial field configuration of the MV model
has strong instabilities and 
the HW entropy is considered to saturate even without showing
the intermediate KS rate.

The HW entropy production rate per degrees of freedom
in the early stage is almost independent of the lattice size.
In addition to $32^3$ and $64^3$ lattices shown in Fig.~\ref{fig:HWE}, 
similar production rate is found on smaller lattices, $4^3$, $8^3$ and $16^3$.
At least, this lattice-size independence dose not come from the chaoticity of the system because the KS rates depend on the lattice size.
The energy density on the $32^3$ lattice is
$\varepsilon=3.53$, and the slopes from 
the local and intermediate KS rates are evaluated as 
$0.08$ and $0.04$, respectively,
which are smaller than the KS rates on the $64^3$ lattice.
This fact suggests that another possible mechanism exists to create the HW entropy such as the initial instability.

The amount of the produced entropy on the $64^3$ lattice is
$\Delta S/N_D \simeq 0.4$ and may be in the same order 
of the expected entropy production.
The longitudinal thickness of glasma at the initial stage should be
in the order of $Q_s^{-1}$,
and the present calculation in the nonexpanding geometry
corresponds to a very thick nuclei, $aL=120\,Q_s^{-1}$.
The produced entropy per unit rapidity for color SU(3) is expected to be
\begin{align}
\frac{\Delta S_\mathrm{HW}}{120 \Delta Y}
=\frac{0.4\times 3 (N_c^2-1) L^3}{1200}
\simeq 2000
\ .
\end{align}
This value is around half of the expected entropy, $\Delta S/\Delta Y\simeq 4500$ \cite{EntropyReview},
but several systematic uncertainties in the present setup
could easily account for a factor of two.
Calculation of the entropy production in an expanding geometry is desired.

\subsection{Classical equilibration}

\begin{figure}[tbhp]
\begin{center}
\includegraphics[width=100mm]{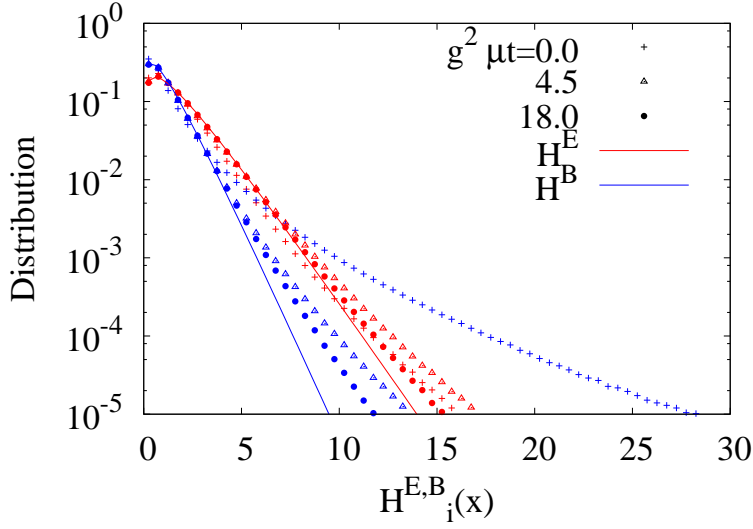}
\caption{Electric and magnetic local energy distribution in the SU(2) Yang-Mills theory on the $64^3$ lattice.
The red (blue) crosses, triangles and circles show the electric (magnetic) energy distributions at $g^2\mu t=0, 4.5$ and $18$, respectively.
The red (blue) line shows a fit function $\sqrt{H^{E(B)}}\exp(-H^{E(B)}/T)$ to the electric (magnetic) energy distribution at $g^2\mu t=18$.}
\label{fig:Edist}
\end{center}
\end{figure}

The present calculation shows that the HW entropy approximately saturates 
at $g^2\mu t \simeq 7 (5)$
on the $32^3$ $(64^3)$ lattice,
then some kind of quasi-stationary state is formed.
In Fig.~\ref{fig:Edist}, we show the distribution
of the electric and magnetic local energies,
\begin{align}
H^E_i(\bold{x})=\frac12 \sum_{a=1}^{N_c^2-1} \left(E^a_i(\bold{x})\right)^2
\ ,\quad
H^B_i(\bold{x})=\frac12 \sum_{a=1}^{N_c^2-1} \left(B^a_i(\bold{x})\right)^2
\quad (i=x,y,z)\ .
\end{align}
In the thermal equilibrium in the classical regime,
the distribution of $(A,E)$ would be described by the Boltzmann distribution,
\begin{align}
\mathcal{Z}=\int D\Gamma\,\exp(-H/T)
=\prod_{i,\bold{x}}\left[
\int \frac{d^3E_i(\bold{x})}{2\pi\hbar g^2}\ e^{-H^E_i(\bold{x})/T}
\right]
\int DA\  e^{-\sum_{i,\bold{x}} H^B_i(\bold{x})/T}
\ .
\end{align}
For the electric energy distribution,
we can rewrite the measure as $d^3E=\sqrt{H^E}\,dH^E\,d\Omega$
with $d\Omega$ being the solid angle in the color space,
and the distribution function can be given as $\sqrt{H^E}\exp(-H^E/T)$.
Actually, the electric energy distribution in the later stage
is described well by this distribution except for the high energy region
as shown by the solid line in Fig.~\ref{fig:Edist}.
The magnetic energy distribution is also found to follow the same function
but with a different temperature.
Similar Boltzmann distribution of the energy is found
in Ref.~\cite{Biro1993}.
Thus the saturation
of the HW entropy seems to be related to the quasi-statinary state,
where approximate equilibrium is reached
among the electric energies 
and among the magnetic energies but with a different temperature.

The saturation time and saturated value of the HW entropy per DOF
decrease with increasing lattice size, as shown in Fig.~\ref{fig:HWE}.
It should be noted that the above quasi-statinary state is, however, different from
the true equilibrium of gluons:
In addition that the electric and magnetic temperatures are different,
the long-term evolution with the classical Yang-Mills equation
does not reach the Bose-Einstein distribution of the high-momentum modes
but reach the classical statistical distribution.
Since the classical statistical distribution in field theories
does not have a well-defined continuum limit,
it is reasonable to find the lattice size dependence 
of the saturation time and saturated value of the HW entropy.

\subsection{Isotropization of pressure}

\begin{figure}[h]
\begin{center}
\includegraphics[width=100mm]{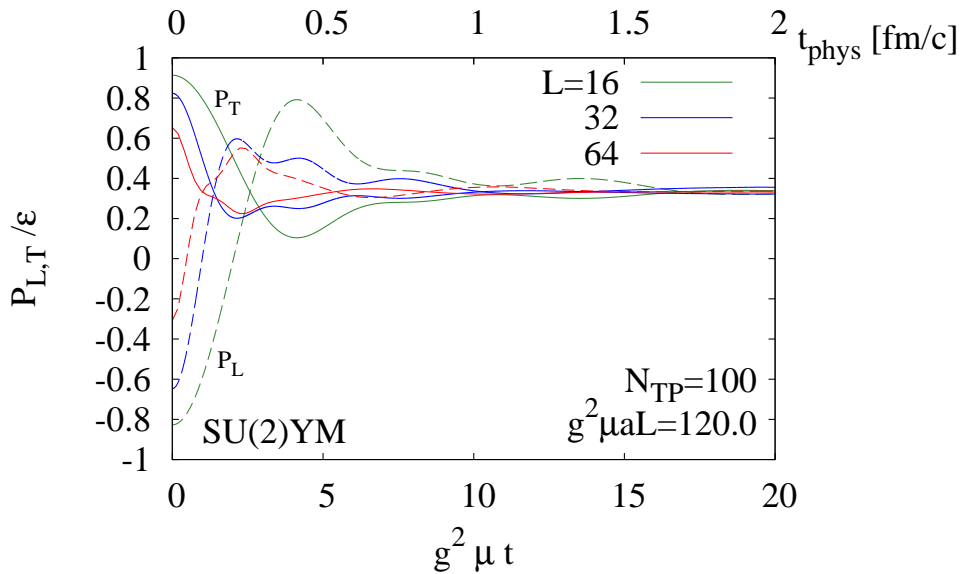}
\caption{Time evolution of the pressure in the SU(2) Yang-Mills (YM) theory.
The horizontal axis is the time axis.
The lower (upper) scale is the dimensionless time scale $g^2 \mu t$ (the physical time $t_\mathrm{phys}$ [fm/c]).
The vertical axis is the longitudinal and transversal pressure normalized by the energy density, $P_{L,T}/\varepsilon$.
The dotted (solid) lines are the longitudinal pressure.
The green, blue and red lines corresponds to the results on the $16^3$, $32^3$ and $64^3$ lattices respectively.}
\label{isoP}
\end{center}
\end{figure}

In Fig.~\ref{isoP},
we show the time evolution of the ratio of the pressure to the energy density ratio
in the longitudinal and transverse directions, $P_{L,T}/\varepsilon$,
on the $16^3$, $32^3$ and $64^3$ lattices.
Because the energy-momentum tensor is traceless, the relation $2 P_L/\varepsilon + P_T/\varepsilon=1$ is satisfied.
While the classical configuration of the MV model has the completely anisotropic pressure $P_L=-P_T$ at initial time,
the quantum fluctuations modifies this relation
and the initial value $P_L/\varepsilon$ ($P_T/\varepsilon$)
is not equal to $1.0$ ($-1.0$).

The isotropization of the pressure can be found to occur in Fig.~\ref{isoP}.
The lattice size dependence of the isotropization time is strong 
in the smaller lattices, $L<32$,
and 
For larger lattices $(L\geq 32)$,
the isotropization time 
almost converges $g^2\mu t\simeq 10$, as seen from the $L=32$ and $L=64$ results,
This isotropization time roughly agrees
with the time of the HW entropy saturation.
It also happens to agree with the isotropization time 
obtained in the expanding geometry
with fluctuation effects from the finite coupling~\cite{Epelbaum13}.

\section{Summary and conclusion}\label{Sec:Summary}

The aim of this paper is to understand the thermalization in the relativistic heavy ion collisions by focusing on the entropy production.
We calculate the Husimi-Wherl (HW) entropy in the Yang-Mills field theory
with the phenomenological initial condition given by the McLerran-Venugopalan (MV) model 
but in the non-expanding geometry.
The HW entropy constructed from the Husimi function plays an important role in thermodynamics of quantum systems and its production implies the thermalization of the system.

We calculate the semiclassical time evolution of the Wigner function by solving the classical equation of motion which keeps the gauge invariance of the HW entropy.
In actual calculations,
we use the product ansatz to reduce the numerical cost
at the cost of breaking the gauge invariance of the HW entropy.
Nevertheless the HW entropy in the product ansatz agrees with the result without the product ansatz within 10-20\% in a few-dimensional quantum mechanical system \cite{Tsukiji2016}
and production rate of the HW entropy in the product ansatz agrees with the Kolmogolov-Sina\"i (KS) rate.

We have found that 
the HW entropy increases linearly in early time and saturates at later times.
The growth rate of the HW entropy is independent of the lattice size and is significantly larger than the intermediate KS rate, defined as the sum of the positive intermediate Lyapunov exponents.
This implies that we cannot explain the entropy production from MV initial condition only by intrinsic chaoticity.
It also suggests that the large amount of the entropy may be produced by the initial instability.
When the HW entropy saturates, the electric and magnetic local energy distributions reach the classical statistical equilibrium
except for the high energy regions.
The saturation time agrees with the equilibrium time of the local energy distribution and the isotropization time of the pressure,
which suggests the thermalization of the gluon field is realized
in the sense of the HW entropy production.
The saturation time is around $1$ fm/c.
In order to reach more quantitative and realistic conclusions,
the evaluation the HW entropy in the expanding geometry is desired,
which is under progress.

\section*{Acknowledgement}
The authors would like to thank Prof. Berndt Muller for useful discussions
and suggestions.
%
This work was supported in part by 
the Grants-in-Aid for Scientific Research from JSPS
 (Nos.
 20540265, 
 23340067, 
 15K05079, 
 15H03663, 
 16K05350, 
and
 16K05365
),
the Grants-in-Aid for Scientific Research on Innovative Areas from MEXT
 (Nos. 
       24105001 and 24105008 
),
and
by the Yukawa International Program for Quark-Hadron Sciences.
T.K. is supported by the Core Stage Back Up program in Kyoto University.

\appendix

\section{Gauge invariance of Husimi-Wehrl entropy}
\label{App:A}

We give proof of the invariance of Husimi-Wehrl entropy with the residual gauge freedom in temporal gauge.

\subsection{Gauge invariance of Wigner function}

In temporal gauge ($A_0=0$),
the gauge transformation is given by
\beq
A_i &\to& A'_i =\Omega A_i \Omega^{-1}+i \Omega \partial_i \Omega^{-1},\nonumber\\
E_i &\to& E'_i =\Omega E_i \Omega^{-1}.\label{gauge}
\eeq

A vector in Hilbert space is transformed by
\beq
|A\rangle \to |A'\rangle=\hat{\Omega}|A\rangle=|\Omega A\Omega^{-1}+i\Omega\partial \Omega^{-1}\rangle.
\eeq

When the density matrix $\rho$ is gauge covariant;
\beq
\rho\to \hat{\rho}'=\hat{\Omega}\hat{\rho}\hat{\Omega}^{-1},\label{DM}
\eeq
we can prove the gauge invariance of the Wigner function.

The Wigner function is transformed by
\begin{align}
\fW[A,E]\to&\fW[A',E']
\nonumber\\
=&\int \frac{Da}{(2\pi\hbar g^2)^{\ND}}\mathrm{e}^{i E'\cdot a/\hbar g^2}\langle A'+a/2|\hat{\rho}'|A'-a/2\rangle\nonumber\\
=&\int \frac{Da'}{(2\pi\hbar g^2)^{\ND}}\mathrm{e}^{i E'\cdot a'/\hbar g^2}\langle A'+a'/2|\hat{\rho}'|A'-a'/2\rangle\nonumber\\
=&\int \frac{Da'}{(2\pi\hbar g^2)^{\ND}}\mathrm{e}^{i E'\cdot (a'-i\Omega\partial \Omega^{-1})/\hbar g^2}\langle A'+\frac{a'}{2}-\frac{i}{2}\Omega\partial \Omega^{-1}|\hat{\rho}'|A'-\frac{a'}{2}+\frac{i}{2}\Omega\partial \Omega^{-1}\rangle\nonumber\\
=&\int \frac{Da}{(2\pi\hbar g^2)^{\ND}}\mathrm{e}^{i \Omega E\Omega^{-1}\cdot \Omega a\Omega^{-1}/\hbar g^2}\langle A+\frac{a}{2}|\hat{\Omega}^{\dagger}\Omega\hat{\rho}\Omega^{-1}\Omega|A-\frac{a}{2}\rangle\nonumber\\
=&\fW[A,E].\label{gaugeinv}
\end{align}
We use the transformation of the $|A\pm a/2\rangle$,
\beq
|A\pm a/2\rangle\to \hat{\Omega}|A\pm a/2\rangle&=&|\Omega A\Omega^{-1} \pm \Omega a\Omega^{-1}/2+ i\Omega\partial \Omega^{-1}\rangle\nonumber\\
&=&|A' \pm a'/2\mp i\Omega\partial \Omega^{-1}/2\rangle.\label{braG}
\eeq

The equation (\ref{gaugeinv}) shows the gauge invariance of the Wigner function.

\subsection{Gauge invariance of Husimi function and Husimi-Wehrl entropy}

It is easy to prove the gauge invariance of the Husimi function from the above discussion.

The gauge transformation of the Husimi function is given by 
\begin{align}
\fH[A,E]\to&\fH[A',E']
\nonumber\\
=&\int \frac{D\bar{A}D\bar{E}}{(\pi\hbar g^2)^{\ND}}\mathrm{e}^{-\Delta(A'-\bar{A})^2/\hbar-(E'-\bar{E})^2/\Delta\hbar g^2}\fW[\bar{A},\bar{E}]\nonumber\\
=&\int \frac{D\bar{A'}D\bar{E'}}{(\pi\hbar g^2)^{\ND}}\mathrm{e}^{-\Delta(A'-\bar{A}')^2/\hbar g^2-(E'-\bar{E}')^2/\Delta\hbar g^2}\fW[\bar{A}',\bar{E}']\nonumber\\
=&\int \frac{D\bar{A}D\bar{E}}{(\pi\hbar g^2)^{\ND}}\mathrm{e}^{-\Delta(\Omega A\Omega^{-1}-\Omega \bar{A}\Omega^{-1})^2/\hbar g^2-(\Omega E\Omega^{-1}-\Omega\bar{E}\Omega^{-1})^2/\Delta\hbar g^2}\fW[\bar{A},\bar{E}]\nonumber\\
=&\fH[A,E].
\end{align}
This equation show the gauge invariance of Husimi function.
The gauge invariance of Husimi-Wehrl entropy follows from these facts.

\subsection{Gauge invariance in semiclassical approximation}

In this subsection, we prove that the semiclassical time evolution dose not break the gauge invariance of the HW entropy.

When the Wigner function is gauge invariant at initial time, 
it is gauge invariant at any time in semiclassical approximation.
\begin{align}
\fW[A,E;t]\to&\fW[A',E';t]
\nonumber\\
=&\fW[A',E';t=0]\nonumber\\
=&\fW[A,E;t=0]=\fW[A,E;t]
\end{align}
Because the classical path is gauge covariant, 
the $(A',E';t)$ at time $t$ and $t=0$ are on the same gauge orbit.

Therefore, the semiclassical time evolution keeps the gauge invariant of the HW entropy.

\section{Lyapunov exponents in SU(2) Yang-Mills theory}
\label{App:B}
In this Appendix, we show the calculated results of the Lyapunov exponents in the SU(2) Yang-Mills theory and show that the Lyapunov exponents are proportional to $\varepsilon^{1/4}$, where $\varepsilon$ is the energy density. Results for the SU(3) Yang-Mills theory is given in Ref. \cite{kunihiro}.
Our numerical formalism and set-up are the same as that of Ref. \cite{kunihiro}.
To detect the intrinsic property of the system such as chaoticity,
we set the initial condition as $E=0$ and $A$ is randomly chosen around zero.

\subsection{Results}
\begin{figure}[tbhp]
\begin{center}
\includegraphics[width=100mm]{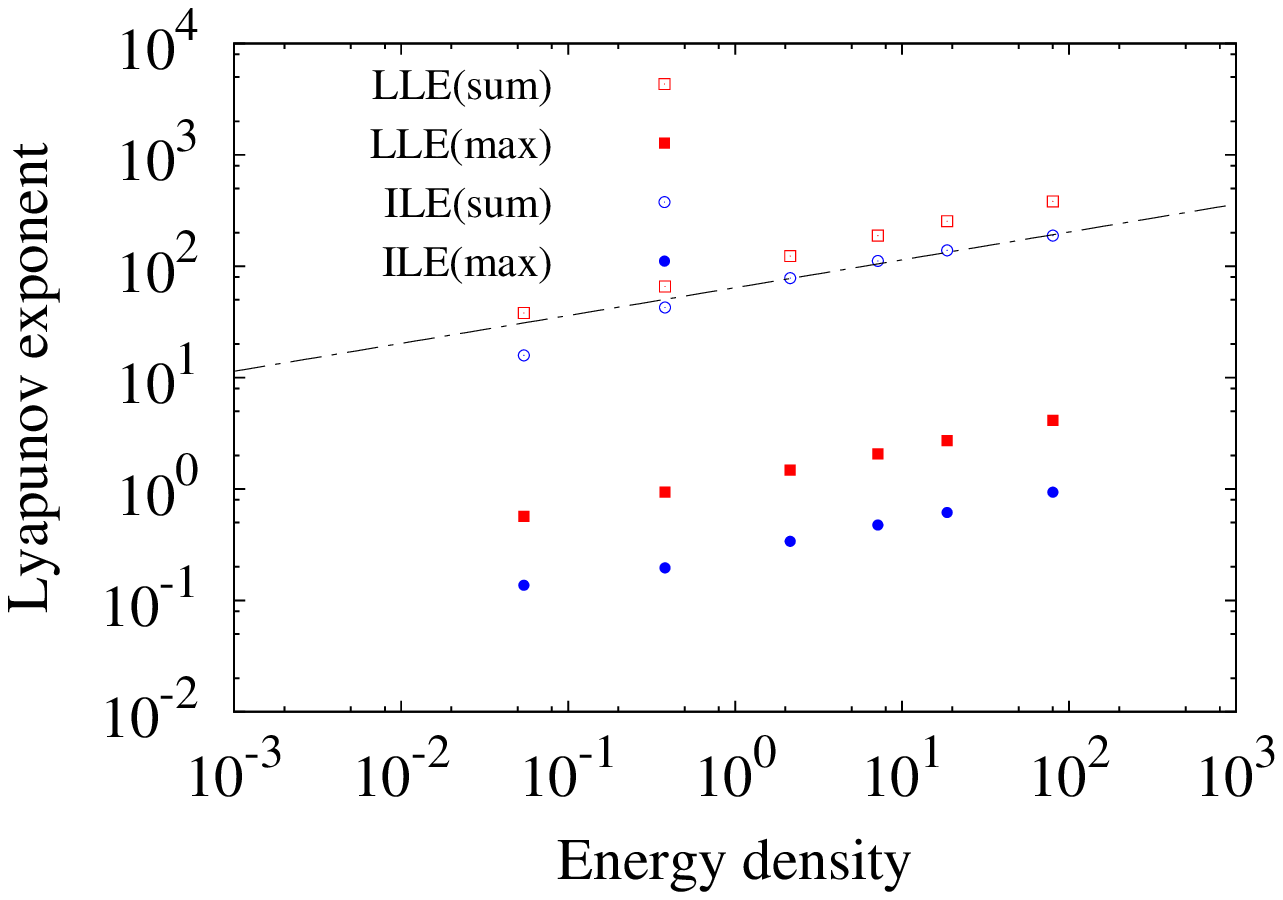}
\caption{Lyapunov exponents in the SU(2) Yang-Mills lattice field theory with random initial condition, $\lambda^{\rm LLE}_{\rm max},\lambda^{\rm LLE}_{\rm KS},\lambda^{\rm ILE}_{\rm max},\lambda^{\rm ILE}_{\rm KS}$. The blocken line is $4^3\times \varepsilon^{1/4}$.}
\label{fig:Lyap}
\end{center}
\end{figure}

We summarize our results in Table \ref{table:Lyap} and Fig. \ref{fig:Lyap}.
Our results show the Lyapunov exponents are proportional to the $\varepsilon^{1/4}$ and we determine the coefficients by fitting the results;
\beq
\lambda^{\rm LLE}_{\rm max}&=&c^{\rm LLE}_{\rm max}\times\varepsilon^{1/4}=1.3\times\varepsilon^{1/4},\\
\lambda^{\rm LLE}_{\rm KS}/L^3&=&c^{\rm LLE}_{\rm KS}\times\varepsilon^{1/4}=1.9\times\varepsilon^{1/4},\\
\lambda^{\rm ILE}_{\rm max}&=&c^{\rm ILE}_{\rm max}\times\varepsilon^{1/4}=0.3\times\varepsilon^{1/4},\\
\lambda^{\rm ILE}_{\rm KS}/L^3&=&c^{\rm ILE}_{\rm KS}\times\varepsilon^{1/4}=1.0\times\varepsilon^{1/4}.
\eeq

\begin{table}[htbp]
\begin{center}
\caption{Lyapunov exponents in SU(2) classical Yang-Mills theory}
  \begin{tabular}{cccccc} \hline \hline
    $L^3$ & $\varepsilon$ & $\lambda^{\rm LLE}_{\rm max}$ & $\lambda^{\rm LLE}_{\rm KS}$ & $\lambda^{\rm ILE}_{\rm max}$ & $\lambda^{\rm ILE}_{\rm KS}$ \\ \hline 
    $4^3$ & 0.054 & 0.569 & 38.1 & 0.137 & 15.9 \\
    $4^3$ & 0.38 & 0.938 & 66.0 & 0.196 & 42.7\\
    $4^3$ & 2.14 & 1.48 & 124 & 0.339 & 78.2\\
    $4^3$ & 7.17 & 2.07 & 254 & 0.616 & 112\\
    $4^3$ & 18.6 & 2.73 & 254 & 0.616 & 139\\
    $4^3$ & 79.9 & 4.14 & 383 & 0.939 & 189\\ \hline\hline
  \end{tabular}
\label{table:Lyap}
\end{center}
\end{table}

\subsection{Comparison with Husimi-Wherl entropy}

\begin{figure}[tbhp]
\begin{center}
\includegraphics[width=100mm]{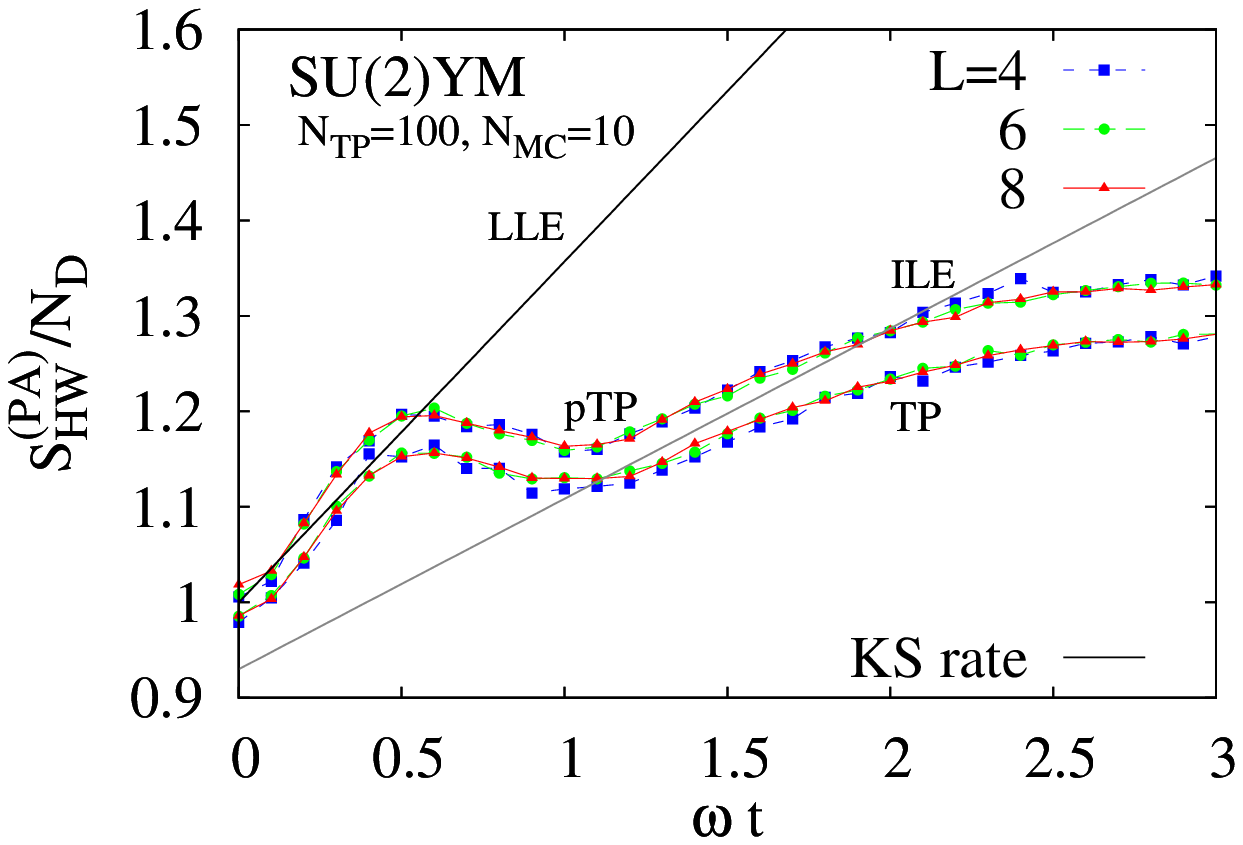}
\caption{The time evolution of HW entropy by TP and pTP methods in the SU(2) Yang-Mills (YM) lattice field theory with random initial condition.
The blue squre, green circle and red triangle lines are the HW entropy per one degree of freedom on $4^3, 6^3$ and $8^3$ lattices, respectively.
The black (gray) solid line shows the growth rate of local (intermediate) KS rate.}
\label{fig:HWErandom}
\end{center}
\end{figure}

We reexamine
 the result in the Ref. \cite{Tsukiji2016} with the Lyapunov exponents in SU(2) Yang-Mills theory.
Fig. \ref{fig:HWErandom} shows the time evolution of the HW entropy in SU(2) Yang-Mills theory with the Gaussian random initial condition around the origin.
The black (gray) straight line shows the HW entropy with the growth rate given by
the local (intermediate) KS rate defined as
the sum of positive LLE (ILE).
The growth rate caused by instabilities in the early time is characterized
by the local KS rate
and the entropy growth rate in the intermediate time caused by chaoticity,
which is the intrinsic property of the Yang-Mills system,
is characterized by the intermediate KS rate.

\end{document}